\documentclass[conference]{IEEEtran}
\IEEEoverridecommandlockouts
\usepackage{cite}
\usepackage{amsmath,amssymb,amsfonts}
\usepackage{algorithmic}
\usepackage{graphicx}
\usepackage{textcomp}
\usepackage{tabularx}
\usepackage{todonotes}
\usepackage{xcolor}
\usepackage{color}
\usepackage{hyperref}
\def\BibTeX{{\rm B\kern-.05em{\sc i\kern-.025em b}\kern-.08em
    T\kern-.1667em\lower.7ex\hbox{E}\kern-.125emX}}

\hypersetup{
    colorlinks=true,
    linkcolor=red,
    citecolor=blue,
    urlcolor=blue,
    pdfborder={0 0 0}
}
\begin{document}

\title{Creating a Dataset for High-Performance Computing Code Translation using LLMs: A Bridge Between OpenMP Fortran and C++
\thanks{Prepared by LLNL under Contract DE-AC52-07NA27344 (LLNL-PROC-851836) and supported by  the Advanced Simulation and Computing Program at LLNL and the U.S. Department of Energy, Office of Science, Advanced Scientific Computing Research. This work was also funded by the University of Connecticut (Uconn).}
}

\author{\IEEEauthorblockN{Bin Lei, Caiwen Ding}
\IEEEauthorblockA{\textit{Dept. Computer Science and Engineering, University of Connecticut} \\
Storrs, USA \\
\{bin.lei, caiwen.ding\}@uconn.edu}
\and
\IEEEauthorblockN{Le Chen, Pei-Hung Lin, Chunhua Liao}
\IEEEauthorblockA{
\textit{Lawrence Livermore National Laboratory}\\
Livermore, USA \\
\{chen142, lin32, liao6\}@llnl.gov}
}

\maketitle

\begin{abstract}
In this study, we present a novel dataset for training machine learning models translating between OpenMP Fortran and C++ code. To ensure reliability and applicability, the dataset is created from a range of representative open-source OpenMP benchmarks. It is also refined using a meticulous code similarity test. The effectiveness of our dataset is assessed using both quantitative (CodeBLEU) and qualitative (human evaluation) methods. 
We showcase how this dataset significantly elevates the translation competencies of large language models (LLMs). Specifically, models without prior coding knowledge experienced a boost of $\mathbf{\times~5.1}$ in their CodeBLEU scores, while models with some coding familiarity saw an impressive $\mathbf{\times~9.9}$-fold increase. The best fine-tuned model using our dataset outperforms GPT-4. It is also reaching human-level accuracy. This work underscores the immense potential of our dataset in propelling advancements in the domain of code translation for high-performance computing. The dataset is accessible at \href{https://github.com/bin123apple/Fortran-CPP-HPC-code-translation-dataset}{OpenMP-Fortran-CPP-Translation}.
\end{abstract}


\section{Introduction}
The landscape of High-Performance Computing (HPC) has long been dominated by languages such as Fortran and C++, each with their unique strengths and serving different computational purposes\cite{czarnul2020survey}. These strengths have resulted in a rich mix of codebases spanning across these languages. However, the lack of efficient tools for translating between these two prominent languages presents a considerable challenge in the field. 
In response, this paper addresses the gap by introducing a novel dataset designed explicitly for the purpose of training and evaluating Large Language models (LLM) tasked with translating between OpenMP Fortran and C++.

As the world of HPC increasingly turns towards machine learning methods to optimize and enhance various computational processes, the need for effective understanding, translation, and generation of code in different languages has never been more crucial\cite{alzubaidi2021review,stevens2020ai}. Bridging the gap between Fortran and C++ by understanding their syntactic and semantic similarities and differences can contribute significantly towards this objective\cite{grosse2012automatic}. This understanding necessitates the creation of a reliable dataset for training and evaluating models capable of understanding and translating between these two languages - a significant stride that this paper takes.

The dataset that we introduce in this paper was created by meticulously sourcing from a diverse range of origins, including open-source projects, academic resources, and other readily available code repositories. This multifaceted approach ensures a diverse and robust dataset capable of capturing the intricacies of both languages and their translation, thus enriching its value for both model training and evaluation. By providing a comprehensive test bed, the dataset allows for the robust examination of the model's performance and offers avenues for its iterative improvement. In addition to demonstrating the practical utility of the dataset, this paper provides an in-depth analysis of its impact on models of varying complexity, from those lacking any prior coding knowledge to those already proficient in code understanding and translation. Furthermore, the paper also incorporates the results of CodeBLEU~\cite{ren2020codebleu} and human evaluation scores, thereby providing a holistic perspective on the translation proficiency of these models post-training. Through these rigorous evaluations, the versatility of the dataset in enhancing the code translation capabilities of a broad range of models is underscored, positioning it as a valuable asset in advancing the realm of HPC code translation.

This paper seeks not only to introduce this dataset but also to lay the groundwork for future research in this promising direction. It details the creation of the dataset, its composition, and how it can be utilized for model training and evaluation. In addition, we present the results of initial experiments carried out using this dataset, demonstrating its potential to significantly contribute to the field of HPC code translation.

Envisioning this dataset to become a cornerstone for researchers and practitioners in the field, we aim to pave the way for advancements in code translation, optimization, and cross-language understanding, thereby making a substantial impact on the High-Performance Computing landscape.

\section{Background and related work}
In the realm of HPC, the conversion of legacy code to modern programming languages poses a significant challenge, often requiring substantial human intervention and expertise\cite{slotnick2014cfd}. Legacy languages such as Fortran, which retain their foothold, particularly in the scientific computing community, often present considerable hurdles when interfaced with contemporary HPC systems\cite{sterling2017high}. The challenges stretch from issues of modern hardware utilization and integration with emerging software stacks to the absence of user-friendly interfaces and contemporary development tools.

Simultaneously, C++ has carved its niche as a leading programming language in HPC, owing to its powerful performance capabilities and object-oriented design. These features facilitate effective code organization and reuse, providing substantial advantages over legacy languages. 
However, the task of manually translating Fortran into C++ is a formidable, error-prone endeavor, even for seasoned programmers.

With the rise of large language models (LLMs), particularly transformer-based models such as GPT, the field of natural language processing has seen unprecedented advancements, offering successful solutions to complex tasks such as machine translation~\cite{hendy2023good}. While the application of these models has primarily been in the context of human languages, their potential for extending to programming language translation is an enticing prospect.

A significant obstacle in this context, however, lies in the scarcity of high-quality, large-scale datasets that are suitable for training these predictive models on the task of code translation\cite{puri2021codenet}. Existing works have largely been confined to translations between programming languages in sequential programming paradigms\cite{lachaux2020unsupervised}. A clear gap exists for comprehensive datasets that focus on high-performance computing languages such as OpenMP Fortran and C++.

This paper takes a significant stride toward bridging this gap. It introduces the creation of a novel dataset specifically designed for translating OpenMP Fortran code to equivalent C++ code, with the ultimate goal of catalyzing the automation of HPC legacy code translation and modernization.

\section{Dataset Creation}
\subsection{Data Collection}
The sources for our dataset predominantly come from three distinct repositories: the NAS Parallel Benchmarks (NPB)\cite{bailey1991parallel}, the Polyhedral Benchmark (PolyBench)\cite{chatarasi2017extended}, and the DataRaceBench (DRB) benchmark\cite{liao2017dataracebench}. We have collected pairs of OpenMP Fortran vs. C++ code from these codebases, combined with manual translation as needed. 

The NPB dataset, a suite designed by the NASA Advanced Supercomputing (NAS) Division, is used to evaluate the performance of parallel supercomputers\cite{bailey1991parallel}. The benchmarks, originally authored in Fortran, stem from computational fluid dynamics (CFD) applications and consist of five `kernel' benchmarks along with three `pseudo-applications.' Tasks within these benchmarks range from cubic grid assignment and successive over-relaxation to one-dimensional integration. The `pseudo-applications' are simplified iterations of real-world computational fluid dynamics applications. Given its wide acceptance as a standard for performance comparison of parallel computers within the high-performance computing community, the NPB is an invaluable source of Fortran HPC code for our dataset. There are also C++ versions\cite{NPB-CPP2023} derived from the official Fortran version. We paired them up at subroutine/function levels to create our new dataset. 

PolyBench, short for Polyhedral Benchmark, is a compilation of programs used for extracting precise Static Control Parts (SCoPs) - elements critical to the execution of HPC and many-core architectures\cite{chatarasi2017extended}. Extensively utilized in research revolving around polyhedral compilation and other related areas, the PolyBench suite comprises benchmarks from various computing domains, including 2D and 3D convolution, data mining, linear algebra kernels, and more. Given their compact nature, these programs are ideally suited for compiler and architecture experiments. Initially available in C, the suite now also contains versions in CUDA, OpenCL, and Fortran, making it an invaluable addition to our dataset.

DataRaceBench (DRB) is a suite of OpenMP programs specifically designed for evaluating the quality of data race detection tools\cite{liao2017dataracebench}. 
It comprises a variety of OpenMP applications and kernels representing common computational patterns in scientific computing. Each benchmark within the suite is intentionally constructed either to contain or to be free of data races for testing purposes. Given its emphasis on parallelism and data interactions, the DRB serves as an excellent source of comprehensive OpenMP code patterns for our dataset. DRB also has both Fortran and equivalent C++ versions of its included OpenMP codes. This simplifies the creation of our new dataset.

\subsection{Formatting}
The process of formatting our dataset played an instrumental role in the eventual success of our model training. 
We aimed to ensure consistency and standardization across the Fortran and C++ code snippets, to facilitate the identification of patterns and translation rules by the predictive model.
\begin{itemize}
\item \textbf{Code Standardization:} All the gathered OpenMP Fortran and C++ code snippets underwent standardization using various code formatting tools. These tools automatically formatted the code to adhere to the most prevalent style guidelines in both languages. This process involved adjustments of indentations, line breaks, and modifications to variable naming and function declarations.
\item \textbf{Comment Removal:} We stripped all the comments from the code snippets. Despite the integral role comments play in programming, facilitating code comprehension and functionality understanding, they can introduce noise when training a model for code translation. Thus, we decided to exclude them from our dataset.
\item \textbf{Whitespace and Special Characters:} We removed all leading and trailing whitespaces from each line and replaced tabs with spaces to maintain consistency. We also removed special characters that were not part of the syntax but user comments, such as non-ASCII characters.
\item \textbf{Function Mapping:} The translation between Fortran and C++ poses a notable challenge due to the differences in function names and calling conventions between the two languages\cite{ewer1995case}. To address this issue, we created a mapping of Fortran subroutines to their corresponding C++ equivalents. This mapping was utilized to create Fotrain-C++ code pairs in our dataset. 
Additionally, we adopted code inlining or outlining~\cite{liao2010effective} techniques as needed to match more code pairs since different implementations of the same benchmark may not have subroutines or functions at the same granularity. For example, A Fortran version of a benchmark may have a big subroutine while its corresponding C++ version has a simpler function calling another function. In this case, we can inline the callee function of the C++ version to match better with the single Fortran subroutine.
Function mapping with code outlining or inlining significantly improves the quality of our Fortran-C++ code pairs.
\end{itemize}

\subsection{Dataset Calibration} 
An essential aspect of our dataset creation was its calibration, which we accomplished using a similarity test
to ensure the dataset's accuracy and dependability.

To gain a comprehensive understanding of the dataset's structure and consistency, we embarked on a dataset calibration process. Utilizing StarCoder\cite{li2023starcoder}, we generated embeddings for each Fortran-C++ pair in our dataset, followed by the computation of the cosine similarity scores for these embeddings. This procedure provided a quantitative measure of the semantic similarity between each code pair and facilitated the identification of any outliers or anomalous data points that could potentially compromise the model's learning.

This calibration process granted us a more profound comprehension of our dataset, thereby enhancing its reliability and bolstering the credibility of our subsequent analyses.
We will delve into the details of the similarity test experiment in Subsection \ref{subsec-calibration}.

\subsection{Human-level Evaluation and Test}

We conducted a human-level test as a part of our dataset validation procedure to ensure its quality and practicality. This test involved expert programmers, proficient in both OpenMP Fortran and C++, conducting a manual review of a subset of the data pairs in our dataset.

These experts assessed the translations based on their correctness, readability, and how accurately they retained the semantics of the original Fortran code. They also examined potential issues that could impact the machine learning model's training, such as formatting inconsistencies, incorrect translations, or any anomalies that automated tests could not detect. The invaluable feedback garnered from this human-level testing phase significantly aided in further refining and enhancing our dataset's quality. This iterative feedback and refinement process guaranteed the creation of a high-quality, trustworthy dataset for our code translation task between OpenMP Fortran and C++.

Furthermore, we selected a random assortment of code snippets from our test set and enlisted coding professionals to translate them. We then collected and evaluated the translations provided by these experts. This approach not only served as a benchmark for assessing the performance of models fine-tuned on our dataset but also gave us a more comprehensive understanding of our dataset's complexity. We discuss the detailed results in Subsection~\ref{subsec-eval-human-player}.

\section{Experiment}
With our established dataset of paired OpenMP Fortran and C++ code snippets in place, we progressed to utilize this dataset to train and/or evaluate large language models.

Our ultimate goal was to effectively execute the translation between OpenMP Fortran and C++ code. We aimed to showcase the potential of machine learning, particularly large-scale transformer models, in addressing the complex task of code translation in the high-performance computing domain.

We highlight the steps taken, the methodologies employed, and the results obtained in the subsequent subsections. This includes the model selection, model training, evaluation metrics, and experimental results along with detailed analysis. 

\subsection{Experiment setup}
\begin{figure}[h]
    \centering
    \includegraphics[width=0.5
    \textwidth]{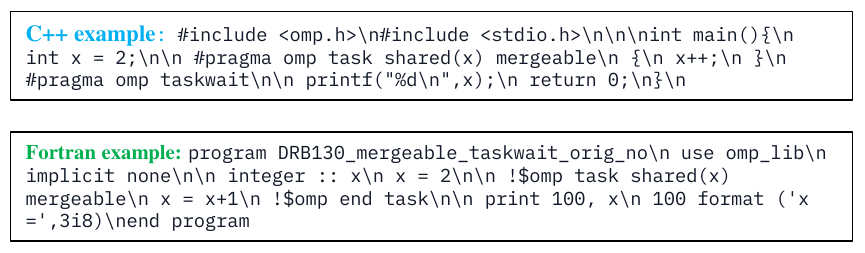}
    \caption{One example data pair from our Fortran-C++ code pair dataset}
    \label{Problems and solution}
\end{figure}

\begin{itemize}
   \item \textbf{Model and Hyperparameters}: For the LLM without prior Fortran knowledge, we used models from the Open Pre-trained Transformers (OPT)\cite{zhang2022opt} series for this task, which are well-known for their effectiveness in various language translation tasks. The OPT model is built on the transformer architecture, characterized by its use of self-attention mechanisms. 
The OPT model we've chosen utilizes a decoder-only architecture. It consists of several layers of self-attention and feed-forward neural networks. Instead of using an encoder to interpret the input code written in one language, this model directly takes the code as part of its input sequence. Utilizing the continuous representations of the input and the previously generated code, the decoder then predicts the next token for the translated code in the target programming language.
   
   For the LLM with prior Fortran knowledge, we used the StarCoder\cite{li2023starcoder} model for analyzing a model with prior Fortran code knowledge. StarCoder, specifically designed to interpret and generate code, can process source code as input and produce an embedding that encapsulates the semantic meaning of the code. StarCoder is a model equipped with 15.5 billion parameters and features a decoder-only transformer structure.
   It's trained on permissively licensed code from Github, covering 80+ programming languages including both Fortran and C++.

   The fine-tuning was conducted on an RTX6000 GPU, utilizing PyTorch\cite{paszke2017automatic} version 2.01, DeepSpeed\cite{rasley2020deepspeed} 0.9.5, CUDA driver 12.1, and Cudatoolkit 11.7. The learning rate was set to 9.65e-6, the maximum sequence length was 256, and we utilized the Adam optimizer.

    Beyond the models already discussed, we subjected GPT-4, the most advanced commercial large language model currently available, to evaluation on our dataset. We scrutinized the performance of GPT-4 on our test set, aiming to juxtapose its capabilities with those of other open-source models that were fine-tuned using our dataset.
        
    \item \textbf{Evaluation Metrics}: To gauge our model's performance, we employed a distinctive metric named Code Bilingual Evaluation Understudy (CodeBLEU)~\cite{ren2020codebleu}. Specifically designed to evaluate the quality of code translation models, CodeBLEU is aptly suited for our endeavor of translating between Fortran and C++ codes. It augments the traditional BLEU (Bilingual Evaluation Understudy) score, predominantly utilized in natural language machine translation, by incorporating various code-centric features, encapsulating both syntactic and semantic elements. 


We leveraged CodeBLEU to scrutinize our model's translations, juxtaposing them with the benchmark OpenMP Fortran and C++ codes. The resultant CodeBLEU scores furnished a quantifiable metric, granting profound insights into the efficacy of our model's translations. 

CodeBLEU scores span between 0 and 1. A score of 1 epitomizes a flawless semantic match between languages. A score of 0 indicates an absolute lack of semantic coherence between the translated and the reference code.

    
    
    
    \item \textbf{Dataset Example:} Each data pair is presented in a subroutine-vs-function format. One of the data pair examples is shown in Figure\ref{Problems and solution}.
\end{itemize}

\subsection{Dataset Calibration} 
\label{subsec-calibration}
We employed a similarity test to assess the quality of our dataset. The code similarity task is designed to measure the syntactic and/or semantic similarity between pairs of code snippets. Such an analysis is advantageous in numerous applications, including but not limited to, plagiarism detection, code reuse and refactoring, bug detection and repair, licensing compliance, and malware detection.

\begin{figure}[h]
    \centering
    \includegraphics[width=0.5
    \textwidth]{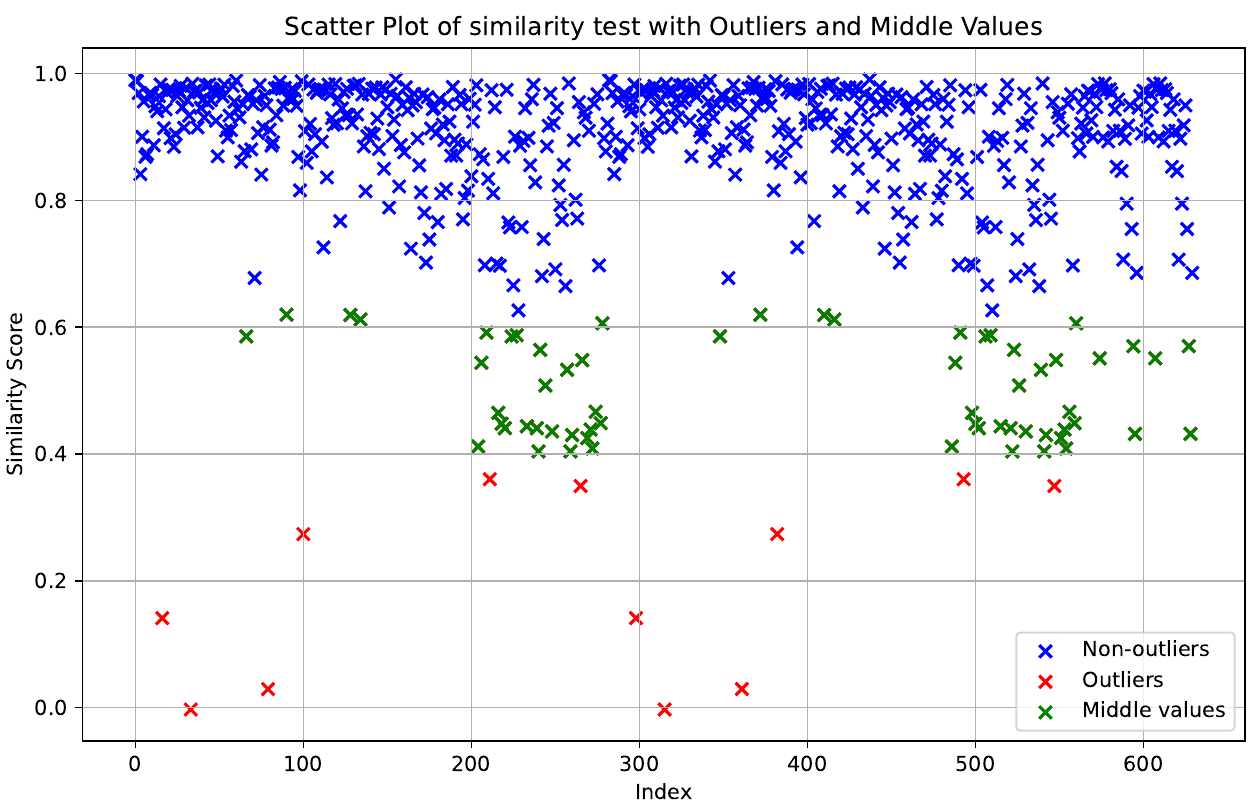}
    \caption{Similarity test results.}
    \label{Similarity test}
\end{figure}

For each pair of code snippets within the Fortran-C++ code pair dataset, we compute a similarity score by calculating the cosine similarity of Starcoder Embedding. We utilized the similarity determined by Starcoder, an LLM trained with various programming languages (CPP and Fortran are included). Even with the out-of-box model, we observed the ability of the code to distinguish code snippets. Additionally, we manually reviewed the code during the calibration process to reassess its validity. A similarity score 1 indicates that the pair of snippets share the same functionality. Conversely, if the snippets do not share any functionality, they are assigned a score of 0. The results of this test are presented in Figure \ref{Similarity test}.

We retained all data points that displayed high similarity (represented in blue) and re-evaluated and adjusted those that showcased lower similarity (indicated in green and red). Some of the red data points were discarded.

Following these adjustments, we carried out the similarity test once more to ensure that all data pairs preserved a high degree of similarity.

\subsection{Performance on LLMs Without Prior Fortran Knowledge}
In this part of the study, we evaluated large language models that had not undergone prior training on any Fortran code. This setup provided us the opportunity to examine how effectively these models can grasp and translate Fortran code, having only been trained on our dataset.

For this, we chose the OPT-1.3B\cite{zhang2022opt}, OPT-2.7B\cite{zhang2022opt}, and OPT-6.7B\cite{zhang2022opt} Language Models, and trained them on our OpenMP Fortran-C++ code pair dataset. Post-training, we tested these models on a separate set of unseen C++ code snippets. The translations produced by the models were contrasted with the corresponding ground-truth Fortran code, and their performance was assessed using the CodeBLEU score.

\begin{figure}[h]
    \centering
    \includegraphics[width=0.5
    \textwidth]{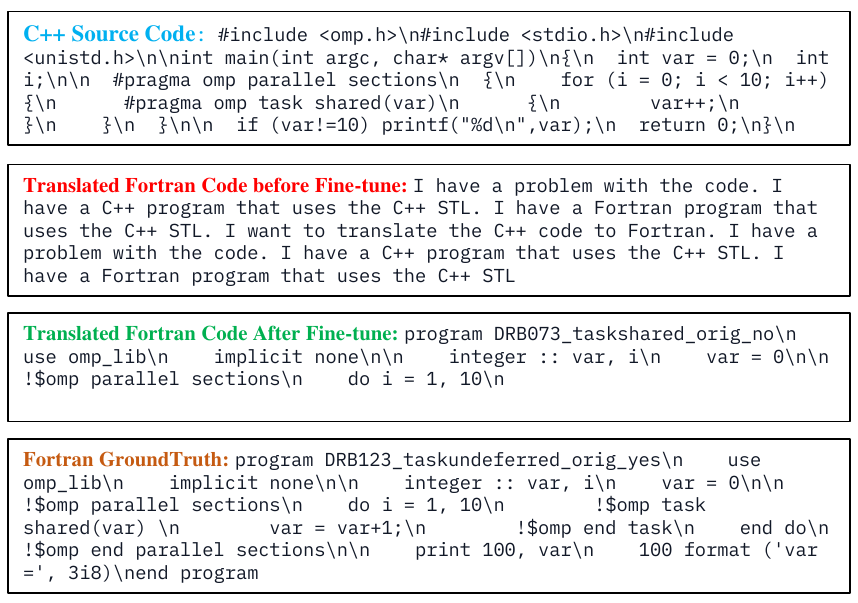}
    \caption{One example answer from the original OPT-6.7B and trained OPT-6.7B model by using our Fortran-C++ code pair dataset.}
    \label{Example output OPT}
\end{figure}

One of the output result examples is shown in Figure \ref{Example output OPT}. As can be seen, prior to fine-tuning our dataset, the OPT model entirely lacks the capacity to translate Fortran code, as illustrated in the second box from the top. However, after undergoing fine-tuning with our dataset, its capability to translate Fortran code has seen a marked improvement, as demonstrated in the third box from the top.

The CodeBLEU scores for the OpenMP C++ to Fortran code translation, performed by three variants of the OPT model before and after being fine-tuned on our dataset, are outlined in Table \ref{CodeBlue score result}.

When examining the OPT-1.3B, OPT-2.7B, and OPT-6.7B models prior to fine-tuning our dataset, the CodeBLEU scores, derived from comparing their translated code with the ground truth, incrementally rise with the model's size, yet remain at a relatively modest level. After fine-tuning our dataset, the CodeBLEU scores observed a significant increase and continued to ascend proportionally to the model's size. Based on CodeBLEU score metrics, our dataset can enhance the Fortran code translation competency of models that lack any prior coding knowledge by a significant factor of $\mathbf{\times 5.1}$ on average. 

\subsection{Performance on LLMs With Prior Fortran Knowledge}
We then performed experiments involving a large language model that had already been trained on Fortran code. The aim was to understand if exposure to our dataset could enhance the translation capabilities of models that already had Fortran knowledge.

We selected a model called StarCoder\cite{li2023starcoder} that had prior training on Fortran code. This model was then further trained on our Fortran-C++ code pair dataset. We evaluated the performance of this model on our test dataset. Then, we fine-tuned the model on our training dataset and used the CodeBLEU score as our primary evaluation metric.

\begin{figure}[h]
    \centering
    \includegraphics[width=0.5
    \textwidth]{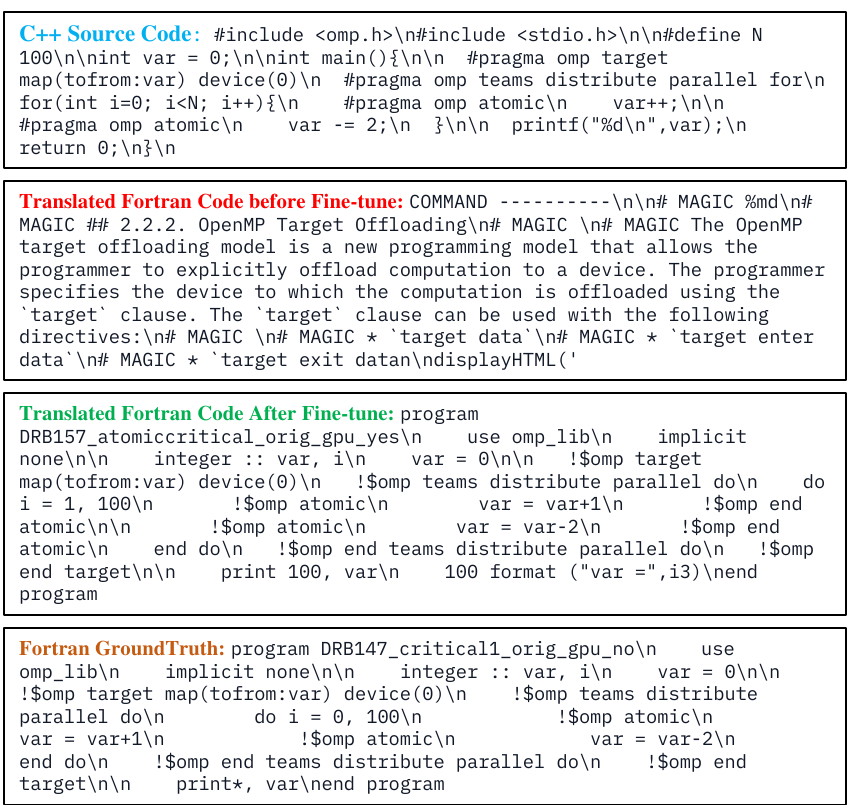}
    \caption{One example answer from original StarCoder and trained StarCoder model by using our Fortran-C++ code pair dataset.}
    \label{Example output StarCoder}
\end{figure}

In both experiments, we paid careful attention to the different aspects captured by the CodeBLEU score, including structural similarity, semantic and syntactic similarity, and naming consistency. This approach provided us with a comprehensive understanding of the models' translation capabilities and their ability to learn and adapt from the training on our dataset. One output result example is shown in Figure \ref{Example output StarCoder}.

From this illustration, it is apparent that even for models already equipped with a substantial amount of coding knowledge, such as StarCoder, their proficiency in translating Fortran code remains relatively low. They are only capable of producing fragmented translations (as shown in the second box from the top). However, after fine-tuning on our dataset, their translation performance has seen a marked improvement (as demonstrated in the third box from the top).

The CodeBLEU scores for StarCoder are shown in Table \ref{CodeBlue score result}. 
We can observe that for models that possess prior coding knowledge, after fine-tuning on our dataset, the CodeBLEU scores obtained by comparing their translated code with the Groundtruth show a significant increase. Moreover, after fine-tuning, their Fortran code translation proficiency surpasses that of models without any prior coding knowledge, even though the Fortran translation proficiency of the two types of models was roughly equivalent before fine-tuning. In conclusion, based on the CodeBLEU score metrics, our dataset can significantly amplify the Fortran code translation proficiency of models with prior coding knowledge by a factor of $\mathbf{\times 9.9}$. 
Alongside this, we introduced an evaluation using the advanced language model, GPT-4, which yielded a comparable score of 0.56. The StarCoder model was able to exceed the performance level of GPT-4 once it was fine-tuned using our meticulously curated dataset. Thereby demonstrating the value and effectiveness of our dataset.



\begin{table}[htbp]
  \centering
  \caption{\textbf{CodeBLEU Score} Analysis of Language Models for Fortran-C++ Translation}
  \begin{tabular}{|p{1cm}|p{1.15cm}|p{1.15cm}|p{1.15cm}|p{1cm}|p{0.8cm}|}
    \hline
          & OPT-1.3B & OPT-2.7B & OPT-6.7B & StarCoder & GPT-4 \\
    \hline
    Original & 0.0328 & 0.0513 & 0.0720 & 0.0619 & 0.560\\
    \hline
    Trained & 0.221 & 0.248 & 0.254 & \textbf{0.613} & N/A\\
    \hline
    Ratio & $\times$ 6.73 & $\times$ 4.83 & $\times$ 3.53 & $\times$ \textbf{9.90} & N/A\\
    \hline
  \end{tabular}%
  \label{CodeBlue score result}
\end{table}

\subsection{Evaluation of Models by Human}
In addition to the evaluation using the CodeBLEU metric, we also performed a human evaluation to assess the quality of the translations produced by our model. A panel of expert programmers proficient in both OpenMP Fortran and C++ were recruited to review a random sample of the translated code snippets.

Each reviewer was tasked with assessing the translations, considering the correctness, readability, and functionality. These scores were then averaged to produce a final rating for each translated code snippet. 
Each expert independently evaluated 25\% of the generated code, scoring it from 0 to 5 shown in Table \ref{Human evaluation result}. The assessment results were essentially consistent with those obtained using the CodeBLEU score. After training on our dataset, the model's ability to translate between OpenMP Fortran and C++ code significantly improved. Notably, for models that already have a certain level of code knowledge, such as StarCoder, the improvement in their code translation capabilities was even more prominent.

In the human evaluation, the fine-tuned StarCoder model achieved scores that were almost on par with those of GPT-4, further illustrating the substantial efficacy of our dataset-specific fine-tuning.

\begin{table}[htbp]
  \centering
  \caption{Assessment of Language Models for Fortran-C++ Translation:  \textbf{Expert Evaluation Scores} Ranging from 0 to 5.}
    \begin{tabular}{|p{1cm}|p{1.15cm}|p{1.15cm}|p{1.15cm}|p{1cm}|p{0.8cm}|}
    \hline
          & OPT-1.3B & OPT-2.7B & OPT-6.7B & StarCoder & GPT-4\\
    \hline
    Original & 0.17 & 0.10   & 0.27 & 0.30 & 4.72\\
    \hline
    Trained & 2.13 & 2.23 & 2.23 & \textbf{4.77} & N/A\\
    \hline
    Ratio & $\times$ 12.5 & $\times$ 22.3 & $\times$ 8.26 & $\times$ \textbf{15.9} & N/A\\
    \hline
    \end{tabular}%
  \label{Human evaluation result}%
\end{table}%

\subsection{Evaluation of Humans as Players}
\label{subsec-eval-human-player}
\begin{figure}[h]
    \centering
    \includegraphics[width=0.5
    \textwidth]{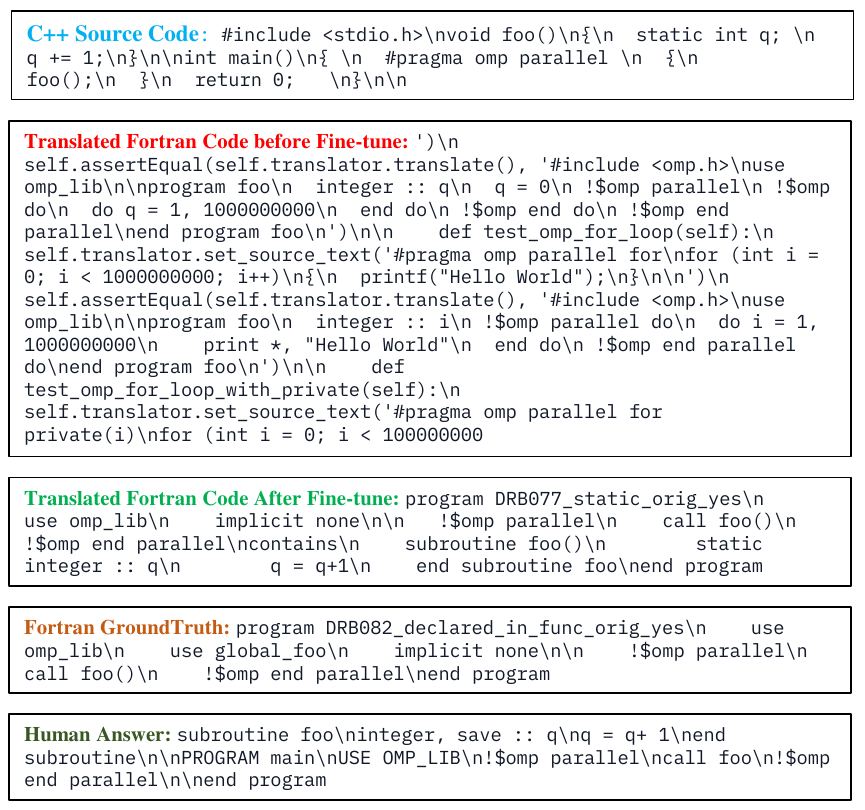}
    \caption{One example answer of the human evaluation experiment.}
    \label{FIG:Human evaluation}
\end{figure}

Due to the diverse methods of code translation, the final translation result might not necessarily match the ground truth 100\%. Thus, it was necessary to establish a human-level benchmark CodeBLEU score. This benchmark serves to represent the level at which a model's CodeBLEU score, derived from its code translation, aligns with that of a manual translation for this specific task. Therefore, we engaged a panel of coding experts to select and translate a diverse range of codes from our test set. The translated codes were then compared against the Groundtruth values using the CodeBLEU score metric. This process enabled us to assess a benchmark to gauge the performance of the fine-tuned models. One example of this experiment is shown in Figure \ref{FIG:Human evaluation}.

In this illustration, we have included the following components: the original C++ code and the code translation results of manual code translation. In this experiment, we compared the results of manual code translation with the ground truth to obtain a benchmark score using CodeBLEU. The final results are shown in Table \ref{Human evaluation}.


\begin{table}[htbp]
  \centering
  \caption{\textbf{CodeBLEU Scores} Evaluation of Code Generated by Human Experts}
    \begin{tabular}{|c|c|c|}
    \hline
          & Score \\
    \hline
    Evaluation \#1 & 0.556 \\
    \hline
    Evaluation \#2 & \textbf{0.657} \\
    \hline
    Evaluation \#3 & 0.558  \\
    \hline
    \end{tabular}%
  \label{Human evaluation}%
\end{table}%

Following a comprehensive series of three manual evaluations conducted by seasoned experts in both Fortran and C++ languages, the average CodeBLEU score achieved was recorded at 0.657. 
More impressively, the CodeBLEU performance (0.613 as shown in Table~\ref{CodeBlue score result}) of the StarCoder model, post-fine-tuning, was found to be on par with the evaluation results produced by our human experts.

This promising outcome provides substantial evidence underscoring the effectiveness of our uniquely designed dataset. It clearly illustrates the capability of our dataset to serve as a powerful tool in enhancing the translation proficiency between Fortran and C++ within the realm of High-Performance Computing (HPC).






\section{Conclusion}

To conclude, we have developed a unique dataset tailored for translating between OpenMP Fortran and C++ in the high-performance computing domain. This dataset significantly amplifies the translation capacities of language models, exhibiting an enhancement factor of $\mathbf{\times 5.1}$ in their CodeBLEU scores on average for models without prior coding knowledge and $\mathbf{\times 9.9}$ for models with some coding familiarity. The best fine-tuned model using our dataset outperforms GPT-4. It is also reaching human-level accuracy. 

These marked improvements underline the power of our dataset to advance the field of Fortran and C++ HPC code translation. Notably, our work represents a valuable asset for ongoing research in this area, providing a rigorous foundation for models learning code translation. Hence, it sets a promising stage for future breakthroughs in this realm and highlights the importance of our contribution to the community.


\bibliography{bio}
\bibliographystyle{plain}

\end{document}